\documentclass{article}

\usepackage{geometry}

\def\ve{\varepsilon}
\def\G{\Gamma}
\def\no{\nonumber}
\def\rar{\rightarrow}
\def\dis{\displaystyle}
\def\le{\left(}
\def\ri{\right)}

\def\d{\partial_{\ve}}
\def\du{\partial_{\ve_1}}
\def\dd{\partial_{\ve_2}}

\def\M{(\omega^{\ve}f(\ve)+f(-\ve))}

\begin{document}

\begin{titlepage}
{\flushleft{DESY~13--067\hfill ISSN 0418-9833}
\flushleft{BI--TP 2013/08}}
\vskip 2cm
\begin{center}
{\Large \bf   Two-fold Mellin-Barnes transforms of Usyukina-Davydychev
functions}\\
\vskip 1cm  
Bernd A. Kniehl $^{(a)},$
Igor Kondrashuk $^{(b,c)},$
Eduardo A. Notte-Cuello $^{(d)},$\\ 
\vskip 2mm
Ivan Parra Ferrada $^{(e)},$
Marko Rojas-Medar $^{(b)}$ \\
\vskip 5mm  
{\it  (a) II. Institut f\"ur Theoretische Physik, Universit\"at Hamburg, \\
          Luruper Chaussee 149, 22761 Hamburg, Germany} \\
{\it  (b) Grupo de Matem\'atica Aplicada, Departamento de Ciencias B\'asicas,  \\ 
         Universidad del B\'\i o-B\'\i o,  Campus Fernando May, 
          Casilla 447, Chill\'an, Chile} \\
{\it  (c) Fakult\"at f\"ur Physik, Universit\"at Bielefeld,
Universit\"atsstra\ss e~25, 33615 Bielefeld, Germany} \\
{\it  (d) Departamento de Matem\'aticas, Facultad de Ciencias, Universidad de La Serena, \\ 
          Av. Cisternas 1200, La Serena, Chile}      \\
{\it  (e) Carrera de Pedagogia en Matem\'atica, Facultad de Educaci\'on y Humanidades, \\
          Universidad del B\'\i o-B\'\i o, Campus Castilla, Casilla 447, 
          Chill\'an, Chile} 
\vskip 6mm
\end{center}

\begin{abstract}
In our previous paper [Nucl.\ Phys.\ B {\bf 870} (2013) 243], we showed that
multi-fold Mellin-Barnes (MB) transforms of Usyukina-Davydychev (UD)
functions may be reduced to two-fold MB transforms.
The MB transforms were written there as polynomials of logarithms of ratios of
squares of the external momenta with certain coefficients.
We also showed that these coefficients have a combinatoric origin.
In this paper, we present an explicit formula for these coefficients.
The procedure of recovering the coefficients is based on taking the
double-uniform limit in certain series of smooth functions of two variables
which is constructed according to a pre-determined iterative way.
The result is obtained by using basic methods of mathematical analysis.
We observe that the finiteness of the limit of this iterative chain of smooth
functions should reflect itself in other mathematical constructions, too,
since it is not related in any way to the explicit form of the MB transforms.
This finite double-uniform limit is represented in terms of a differential
operator with respect to an auxiliary parameter which acts on the integrand of
a certain two-fold MB integral. 
To demonstrate that our result is compatible with original representations of
UD functions, we reproduce the integrands of these original integral
representations by applying this differential operator to the integrand of the 
simple integral representation of the scalar triangle four-dimensional integral
$J(1,1,1-\ve).$ 
\vskip 0.5 cm
\noindent Keywords: Bethe-Salpeter equation; Mellin-Barnes transform;
Usyukina-Davydychev functions
\vskip 0.5 cm
\noindent PACS: 02.30.Gp, 02.30.Nw, 02.30.Uu, 11.10.St 
\end{abstract}
\end{titlepage}

\section{Introduction}

The Mellin-Barnes transformation is an efficient method for the calculation of
Feynman diagrams \cite{Smirnov,Boos:1990rg,Davydychev:1992xr}. 
This method has been playing an important role in multi-loop calculations
within maximally supersymmetric Yang-Mills theory
\cite{Bern:2005iz,DelDuca:2009au}, in which a certain class of master
integrals contributing to the Feynman diagrams is reduced to integrals
corresponding to scalar ladder diagrams \cite{Bern:2005iz}, at least through
three loops in momentum space. 

The scalar ladder diagrams at any loop order in $d=4$ space-time dimensions
were studied for the first time in
Refs.~\cite{Belokurov:1983km,Usyukina:1992jd,Usyukina:1993ch} in momentum
space. 
The calculation of the momentum integrals results in UD functions
\cite{Usyukina:1992jd,Usyukina:1993ch}.
Their MB transforms were investigated in
Refs.~\cite{Allendes:2009bd,Allendes:2012mr}. 
These functions possess remarkable properties, in particular they are invariant
with respect to Fourier transformation
\cite{Kondrashuk:2008ec,Kondrashuk:2008xq}.
Later, this property was generalized via the MB transform to any three-point
Green's function in the massless theory for arbitrary space-time
dimension \cite{Allendes:2009bd,Kondrashuk:2009us}.   
Due to this invariance with respect to Fourier transformation, UD functions
appear in the results of calculations of the Green's functions in position
space 
\cite{Cvetic:2004kx,Cvetic:2006iu,Cvetic:2007fp,Cvetic:2007ds,Mitra:2008yr,Mitra:2008pw,Mitra:2009zm}.  

In Ref.~\cite{Allendes:2012mr},  multi-fold MB transforms of UD functions were
reduced to two-fold MB transforms.
This result allows us to simplify the analysis of the recursive property of MB
transforms of UD functions to the analysis of the recursive property of the
smooth functions that appear in the integrands of MB transforms
\cite{Allendes:2012mr}.   
The MB transform of the UD function with number $n$ turns out to be a linear
combination of three MB transforms of the UD function with number $n-1$, where
each of these three MB transforms depends on two independent variables
$\ve_1,\ve_2$ in a proper, well-defined manner.
The coefficients in front of these MB transforms with lower indices are
singular in these two independent variables in the limit in which these
variables vanish.
However, these singularities cancel each other, and the double-uniform limit
always exists and is finite for each number $n$.  

This limit is a sum of powers of logarithms of certain arguments multiplied by
derivatives of the Euler $\G$ function constructed in a such way that the sum
of the power of the logarithm and the order of the derivative is a fixed number
which depends on the number of the corresponding UD function which in turn
coincides with the number of the rungs in the given ladder diagram.
In Ref.~\cite{Allendes:2012mr}, we constructed the recursive procedure, but did
not consider the MB transforms of the higher UD functions.
In the present paper, we find the explicit form of the MB transforms of the
higher UD functions by establishing the coefficients in front of the powers of
logarithms.
The arguments of the logarithms are ratios of squares of the external incoming
momenta of the ladder diagrams.
As we pointed out in Ref.~\cite{Allendes:2012mr}, the coefficients have their
origin in combinatorics and are certain combinations of the combinatorial
numbers $C_m^n.$ 

These coefficients appear in the result of the calculation of the
double-uniform limit.
In the present paper, we obtain the result for this limit in terms of a
differential operator with respect to an auxiliary parameter $\ve$ which acts
on the integrand of a certain two-fold MB integral.
This MB integral corresponds to the MB transform of the triangle scalar
integral $J(1,1,1-\ve)$ of Ref.~\cite{Usyukina:1992jd} in $d=4$ dimensions.
This scalar integral has a simple integral representation, which is given in
Ref.~\cite{Usyukina:1993ch}.
Taking the limit $\ve \rar 0$ in this original simple integral representation
of $J(1,1,1-\ve)$ produces the first UD function in
Ref.~\cite{Usyukina:1992jd}.
The UD functions with higher number $n$ of rungs have simple integral
representations in the form of integrals over just one variable, too
\cite{Usyukina:1993ch}. 
The integrands in these representations were found in the form of polynomials
of logarithms of the variable of integration.
We reproduce these integrands by applying the differential operator
constructed in the present paper to the integrand of $J(1,1,1-\ve)$.
We also discuss the relation of our approach to the representation of the
ladder diagrams obtained in terms of single-valued harmonic polylogarithms in
Ref.~\cite{Drummond:2013nda}. 

\section{Recursive relations for MB transforms}

In Ref.~\cite{Allendes:2012mr}, the MB transform of the second UD function was
found in terms of the double MB transform,   
\begin{eqnarray} \label{M2}
\oint_C du~dv~x^u~y^v M_2^{(u,v)}(\ve_1,\ve_2,\ve_3) = \no\\
\frac{J}{2}\oint_C du~dv~x^u~y^v \left[ \frac{a(\ve_1)}{\ve_2\ve_3} y^{\ve_2} \left[ x^{\ve_1}\Gamma(-u-\ve_1)\Gamma(-v+\ve_1) 
+ y^{\ve_1}\Gamma(-u+\ve_1)\Gamma(-v-\ve_1) \right]  \right. \no\\
\left. 
+ \frac{a(\ve_3)}{\ve_1\ve_2} \left[ x^{-\ve_3}\Gamma(-u+\ve_3)\Gamma(-v-\ve_3) 
+ y^{-\ve_3}\Gamma(-u-\ve_3)\Gamma(-v+\ve_3) \right] \right. \no\\
\left. 
+ \frac{a(\ve_2)}{\ve_1\ve_3} x^{\ve_1} \left[ x^{\ve_2}\Gamma(-u-\ve_2)\Gamma(-v+\ve_2) 
+ y^{\ve_2}\Gamma(-u+\ve_2)\Gamma(-v-\ve_2) \right]\right] \times \no\\
\times\G(-u)\G(-v) \G^2(1+u+v),  
\end{eqnarray}
where the definition 
\begin{eqnarray}
a(\ve) = \left[ \G(1-\ve)\G(1+\ve)\right]^{-1},  ~~ a_0^{(n)} = (a(\ve))^{(n)}_{\ve=0} 
\end{eqnarray}
has been introduced.
In the limits of vanishing $\ve_i$, which are always subject to the condition 
\begin{eqnarray}
\ve_1 + \ve_2 + \ve_3 = 0,
\end{eqnarray}
we may write  
\begin{eqnarray} \label{limit-M2}
\lim_{\ve_2 \rar 0, \ve_1 \rar 0} \oint_C du~dv~x^u~y^v M_2^{(u,v)}(\ve_1,\ve_2,\ve_3) = \no\\
\oint_C du~dv~x^u~y^v \G(-u)\G(-v) \G^2(1+u+v)\left[\frac{3}{2}\le a(\ve)\G(-u-\ve)\G(-v + \ve)\ri^{(2)}_0 + \right. \no\\
\left. \frac{3}{2}\ln\frac{x}{y}\le a(\ve)\G(-u-\ve)\G(-v + \ve)\ri'_0 + \frac{1}{4}\ln^2\frac{x}{y}\G(-u)\G(-v)\right].
\end{eqnarray}
As we can see, a finite limit exists.
This is to be expected, since this expression was constructed from another one
for which a finite limit exists. 

To be more concise, we introduce another notation, namely 
\begin{eqnarray}
M_1^{(u,v)}(\ve) \equiv \frac{1}{2}\left[x^{\ve}a(\ve)\Gamma(-u-\ve)\Gamma(-v+\ve)  + y^{\ve}a(-\ve)\Gamma(-u+\ve)\Gamma(-v-\ve)\right]
\nonumber\\
\times\G(-u)\G(-v) \G^2(1+u+v).
\end{eqnarray}
With this notation, we write instead of the previous integral relation the
following one:
\begin{eqnarray}
\oint_C du~dv~x^u~y^v M_2^{(u,v)}(\ve_1,\ve_2,\ve_3) = \no\\
J\oint_C du~dv~x^u~y^v \left[\frac{1}{\ve_2\ve_3} y^{\ve_2} M_1^{(u,v)}(\ve_1) + \frac{1}{\ve_1\ve_2} M_1^{(u,v)}(-\ve_3) + \frac{1}{\ve_1\ve_3} x^{\ve_1} M_1^{(u,v)}(\ve_2) \right]. 
\end{eqnarray}

The formula of Ref.~\cite{Allendes:2012mr} relating the MB transforms of the
third and the second UD functions reads:
\begin{eqnarray} \label{M3}
\oint_C du~dv~x^u~y^v M_3^{(u,v)}(\ve_1,\ve_2,\ve_3) = \no\\
\oint_C du~dv~x^u~y^v \left[\frac{1}{\ve_1\ve_2} x^{-\ve_1} y^{-\ve_2} M_2^{(u,v)}(\ve_1,\ve_2,\ve_3) + \frac{J}{\ve_2\ve_3} x^{-\ve_1} M_2^{(u,v)}(\ve_1) 
+  \frac{J}{\ve_1\ve_3} y^{-\ve_2} M_2^{(u,v)}(\ve_2)\right], 
\end{eqnarray}
in which we use the definitions of Ref.~\cite{Allendes:2012mr}.
The limit of vanishing $\ve_i$ is finite and given by 
\begin{eqnarray}
\lim_{\ve_2 \rar 0, \ve_1 \rar 0}  \oint_C du~dv~x^u~y^v M_3^{(u,v)}(\ve_1,\ve_2,\ve_3) = \no\\
\oint_C du~dv~x^u~y^v \G(-u)\G(-v) \G^2(1+u+v)\left[\frac{5}{12} \le a(\ve)\G(-u-\ve)\G(-v + \ve)\ri^{(4)}_0 + \right. \no\\
\left.
\frac{5}{6}\ln \frac{x}{y} \le a(\ve)\G(-u-\ve)\G(-v + \ve)\ri^{(3)}_0 + 
\frac{1}{2}\ln^2 \frac{x}{y} \le a(\ve)\G(-u-\ve)\G(-v + \ve) \ri^{(2)}_0 \right. \no\\
\left.
+ \frac{1}{12}\ln^3 \frac{x}{y} \le a(\ve)\G(-u-\ve)\G(-v + \ve)\ri'_0  \right]. \label{limit-M3}
\end{eqnarray}
This limit should be taken after substituting the expression for $M_2$ in
Eq.~(\ref{M2}) into the expression for $M_3$ in Eq.~(\ref{M3}). 
The coefficient $J$ is defined in Ref.~\cite{Allendes:2012mr} as 
\begin{equation} \label{J}
J= \frac{\Gamma(1-\varepsilon_1)\Gamma(1-\varepsilon_2)\Gamma(1-\varepsilon_3)}{\Gamma(1+\varepsilon_1)\Gamma(1+\varepsilon_2)\Gamma(1+\varepsilon_3)}.
\end{equation}

According to formulae given in Section~4.4 of Ref.~\cite{Allendes:2012mr}, we
have the following expression for the MB transform $M_4$ of the fourth UD
function in terms of the MB transform $M_3$ of the third UD function: 
\begin{eqnarray} \label{M4}
\oint_C du~dv~x^u~y^v M_4^{(u,v)}(\ve_1,\ve_2,\ve_3) = \oint_C du~dv~x^u~y^v \frac{J}{\ve_2\ve_3}M_3^{(u,v)}(\ve_1) \no\\
+ \oint_C du~dv~x^u~y^v\frac{1}{\ve_1\ve_2}M_3^{(u,v)}(\ve_1,\ve_2,\ve_3) + \oint_C du~dv~x^u~y^v\frac{J}{\ve_1\ve_3}M_3^{(u,v)}(\ve_2) =  \no\\
 \oint_C du~dv~x^u~y^v \left[\frac{J}{\ve_2\ve_3}M_3^{(u,v)}(\ve_1) + \frac{1}{\ve_1\ve_2}M_3^{(u,v)}(\ve_1,\ve_2,\ve_3) +  
\frac{J}{\ve_1\ve_3}M_3^{(u,v)}(\ve_2) \right]. 
\end{eqnarray}
Proceeding according to the construction described in
Ref.~\cite{Allendes:2012mr} for the higher UD functions with number $n > 4$,
we may write  
\begin{eqnarray} \label{Mn}
\oint_C du~dv~x^u~y^v M_n^{(u,v)}(\ve_1,\ve_2,\ve_3) = \oint_C du~dv~x^u~y^v \frac{J}{\ve_2\ve_3}M_{n-1}^{(u,v)}(\ve_1) \no\\
+ \oint_C du~dv~x^u~y^v\frac{1}{\ve_1\ve_2}M_{n-1}^{(u,v)}(\ve_1,\ve_2,\ve_3) 
+ \oint_C du~dv~x^u~y^v\frac{J}{\ve_1\ve_3}M_{n-1}^{(u,v)}(\ve_2) \no\\
= \oint_C du~dv~x^u~y^v \left[\frac{J}{\ve_2\ve_3}M_{n-1}^{(u,v)}(\ve_1) + \frac{1}{\ve_1\ve_2}M_{n-1}^{(u,v)}(\ve_1,\ve_2,\ve_3) 
+ \frac{J}{\ve_1\ve_3}M_{n-1}^{(u,v)}(\ve_2) \right].
\end{eqnarray}

In order to simplify the presentation in the previous ladder construction for
higher UD functions, we define  
\begin{eqnarray}\label{f}
f(\ve)= \frac{1}{2}a(\ve)\G(-u-\ve)\G(-v+\ve)\G(-u)\G(-v) \G^2(1+u+v).
\end{eqnarray}

Since the contours of integration in the MB transforms of the UD functions in
Eqs.~(\ref{M2}), (\ref{M3}), and (\ref{Mn}) pass between the leftmost of the
right poles and the rightmost of the left poles in the planes of complex
variables of integration $u$ and $v$, we may work with the limits in
Eqs.~(\ref{limit-M2}) and (\ref{limit-M3}) at the integrand level.
The dependence on the integration variables $u$ and $v$ may be omitted to
simplify the analysis, since it follows from the dependence of $M_1$ on $u$
and $v$ for the higher number of $n$ in $M_n$. 
Due to this observation, it is convenient to introduce the new notation 
\begin{eqnarray}
M_n^{(u,v)}(\ve_1,\ve_2,\ve_3) \equiv \Delta_n (\ve_1,\ve_2,\ve_3).
\end{eqnarray}

We may analyze the functions $\Delta_n (\ve_1,\ve_2,\ve_3)$ as certain
functions of three variables and represent the ladder relations in
Eqs.~(\ref{M4}) and (\ref{Mn}) in the form
\begin{eqnarray}\label{Delta-1-4} 
\Delta_1 (\ve) = x^{\ve}f(\ve)  + y^{\ve}f(-\ve) \no\\
\Delta_2 (\ve_1,\ve_2,\ve_3) = J\left[ \frac{1}{\ve_2\ve_3} y^{\ve_2}\Delta_1 (\ve_1)  + \frac{1}{\ve_1\ve_2} \Delta_1(-\ve_3) + \frac{1}{\ve_1\ve_3} x^{\ve_1} \Delta_1(\ve_2) \right]\no\\
\Delta_3 (\ve_1,\ve_2,\ve_3) =  \frac{1}{\ve_1\ve_2} y^{-\ve_2}x^{-\ve_1} \Delta_2 (\ve_1,\ve_2,\ve_3)  + \frac{J}{\ve_2\ve_3} x^{-\ve_1} \Delta_2(\ve_1) 
+ \frac{J}{\ve_1\ve_3} y^{-\ve_2} \Delta_2(\ve_2) \no\\
\Delta_4 (\ve_1,\ve_2,\ve_3) = \frac{J}{\ve_2\ve_3} \Delta_3(\ve_1) + \frac{1}{\ve_1\ve_2} \Delta_3 (\ve_1,\ve_2,\ve_3)  + \frac{J}{\ve_1\ve_3} \Delta_3(\ve_2).
\end{eqnarray}
For arbitrary number $n>4$, we may write 
\begin{eqnarray} 
\Delta_n (\ve_1,\ve_2,\ve_3) = \frac{J}{\ve_2\ve_3} \Delta_{n-1}(\ve_1) + \frac{1}{\ve_1\ve_2} \Delta_{n-1} (\ve_1,\ve_2,\ve_3)  + \frac{J}{\ve_1\ve_3} \Delta_{n-1}(\ve_2).
\end{eqnarray}
In the next section, we calculate the values $\Delta_n (0)$ of the finite
double-uniform limit
\begin{eqnarray} 
\Delta_n(0) = \lim_{\ve_1 \rar 0,\ve_2 \rar 0}\Delta_n (\ve_1,\ve_2,\ve_3).
\end{eqnarray}
These values correspond to the representations of the MB transforms of UD
functions described in Section~7 of Ref.~\cite{Allendes:2012mr}.

\section{$\Delta_n(0)$ in terms of differential operator}

According to Eq.~(\ref{Delta-1-4}), the expression for $\Delta_2$ may be
explicitly written as    
\begin{eqnarray}
J^{-1}\Delta_2(\ve_1,\ve_2,\ve_3) = \frac{1}{\ve_2\ve_3} y^{\ve_2} \left[ x^{\ve_1}f(\ve_1)
+ y^{\ve_1}f(-\ve_1) \right]
+ \frac{1}{\ve_1\ve_2} \left[ x^{-\ve_3}f(-\ve_3)
+ y^{-\ve_3}f(\ve_3) \right] \no\\
+ \frac{1}{\ve_1\ve_3} x^{\ve_1} \left[ x^{\ve_2}f(\ve_2)
+ y^{\ve_2}f(-\ve_2) \right].
\end{eqnarray}
In order to simplify the analysis, we introduce the notation
$\tilde{\Delta}_2(\ve_1,\ve_2,\ve_3) \equiv y^{-\ve_2}x^{-\ve_1}\Delta_2(\ve_1,\ve_2,\ve_3)$.
For this quantity, we may write
\begin{eqnarray}
 J^{-1}\tilde{\Delta}_2(\ve_1,\ve_2,\ve_3) = \frac{1}{\ve_2\ve_3} \left[f(\ve_1) + \omega^{-\ve_1}f(-\ve_1) \right]
+ \frac{1}{\ve_1\ve_2} \left[ \omega^{\ve_2}f(\ve_1+\ve_2) + \omega^{-\ve_1}f(-\ve_1-\ve_2) \right] \no\\
+ \frac{1}{\ve_1\ve_3} \left[ \omega^{\ve_2}f(\ve_2) + f(-\ve_2) \right],
\end{eqnarray}
where we have defined $\omega \equiv x/y$.
For future use, it is more convenient to represent
$\tilde{\Delta}_2(\ve_1,\ve_2,\ve_3)$ in the form
\begin{eqnarray}
 J^{-1}\tilde{\Delta}_2(\ve_1,\ve_2,\ve_3) = \frac{\omega^{-\ve_1}}{\ve_1}\frac{  \omega^{\ve_1+\ve_2}f(\ve_1+\ve_2) - \omega^{\ve_1}f(\ve_1)}{\ve_2} 
+ \frac{\omega^{-\ve_1}}{\ve_1}\frac{ f(-\ve_1-\ve_2) - f(-\ve_1)}{\ve_2}
\nonumber\\ 
+ \frac{1}{\ve_1(\ve_1+\ve_2)}\le \omega^{-\ve_1}f(-\ve_1) + f(\ve_1)  - \omega^{\ve_2}f(\ve_2) - f(-\ve_2) \ri.
\end{eqnarray}
We may take the limit with respect to the second variable $\ve_2$, and the
result is 
\begin{eqnarray} \label{Delta-2-limit-2}
\tilde{\Delta}_2(\ve_1) = \lim_{\ve_2 \rar 0}   \tilde{\Delta}_2(\ve_1,\ve_2,\ve_3)  = \frac{1}{\ve_1} \omega^{-\ve_1} \le \omega^{\ve_1}f(\ve_1) + f(-\ve_1) \ri'  \no\\
+ \frac{1}{\ve_1^2} \left(f(\ve_1) + \omega^{-\ve_1}f(-\ve_1) - 2f(0) \right).
\end{eqnarray}
The quantity 
\begin{eqnarray} \label{Delta-2-limit}
\Delta_2(0) =    \tilde{\Delta}_2(0) = \lim_{\ve_2 \rar 0, \ve_1 \rar 0}  \tilde{\Delta}_2(\ve_1,\ve_2,\ve_3) =  \frac{1}{2}\ln^2\omega f(0) +3\ln\omega f^{(1)}(0) + 3 f^{(2)}(0) 
\end{eqnarray}
may be written in the alternative form 
\begin{eqnarray} \label{Delta-2-derivative}
\Delta_2(0) =   \frac{1}{2}\left[2f^{(2)}(0)+2\ln\omega f^{(1)}(0)+\ln^2\omega f(0)\right] +  2f^{(2)}(0)+2\ln\omega f^{(1)}(0) \no\\
=  \frac{1}{2}\left[f^{(2)}(\ve)+\omega^{-\ve}\ln^2\omega f(-\ve)+2\omega^{-\ve}\ln\omega f^{(1)}(-\ve)+\omega^{-\ve}f^{(2)}(-\ve) \right]_{\ve=0}\no\\
+  \left[\ln\omega f^{(1)}(\ve)+f^{(2)}(\ve)+\omega^{-\ve}\ln\omega f^{(1)}(-\ve)+\omega^{-\ve}f^{(2)}(-\ve)\right]_{\ve=0}\no\\
=  \frac{1}{2}\left[\d^2\omega^{-\ve}\M\right]_{\ve=0} +  \left[\d \omega^{-\ve}\d\M\right]_{\ve=0}\no\\
=  \frac{1}{2}\left[\d[\d \omega^{-\ve}+2\omega^{-\ve}\d]\M\right]_{\ve=0}.
\end{eqnarray}                                           
Neither this formula nor Eq.~(\ref{Delta-2-limit}) play any important role in
the further construction.
However, the representation in Eq.~(\ref{Delta-2-derivative}) is necessary to
observe that this is a particular case of the general formula for an arbitrary
number $n$ of $\Delta_n(0)$.

The next step is to perform the following operation:
\begin{eqnarray} \label{lemma1}
\lim_{\ve_2 \rar 0}   \dd J^{-1}\tilde{\Delta}_2(\ve_1,\ve_2,\ve_3)  =  
\frac{1}{2\ve_1}\omega^{-\ve_1} \du^2 (\omega^{\ve_1}f(\ve_1)+f(-\ve_1)) \no\\
- \frac{1}{\ve_1^3}(\omega^{-\ve_1}f(-\ve_1)+f(\ve_1)-2f(0)) - \frac{1}{\ve_1^2}\lim_{\ve_2 \rar 0}\dd(\omega^{\ve_2}f(\ve_2)+f(-\ve_2)).
\end{eqnarray}
As may be seen from Eq.~(\ref{Delta-2-limit}), the value
$\tilde{\Delta}_2(\ve_1,\ve_2,\ve_3)$ does not have any singularity in the
variables $\ve_1$ and $\ve_2$.
The same statement is true for its derivative with respect to the variable
$\ve_2$. 

The operation in Eq.~(\ref{lemma1}) is necessary to evaluate the following
term in the chain of functions ${\Delta}_n(0)$.
Indeed, for ${\Delta}_3$, we may write
\begin{eqnarray}
\Delta_3(\ve_1,\ve_2,\ve_3)  =  \frac{1}{\ve_1\ve_2} \le \tilde{\Delta}_2(\ve_1,\ve_2,\ve_3) - J\tilde{\Delta}_2(\ve_1)\ri 
+ \frac{1}{\ve_1(\ve_1+\ve_2)}\le J\tilde{\Delta}_2(\ve_1) - J\tilde{\Delta}_2(\ve_2)\ri.
\end{eqnarray}
Taking into account Eq.~(\ref{lemma1}), we may write 
\begin{eqnarray} \label{Delta-3-limit-2}
\Delta_3(\ve_1) = \lim_{\ve_2 \rar 0}J^{-1} \Delta_3(\ve_1,\ve_2,\ve_3)  =  \frac{1}{2\ve_1^2}\omega^{-\ve_1} \du^2 (\omega^{\ve_1}f(\ve_1)+f(-\ve_1)) \no\\
- \frac{1}{\ve_1^4}(\omega^{-\ve_1}f(-\ve_1)+f(\ve_1)-2f(0)) - \frac{1}{\ve_1^3}\lim_{\ve_2 \rar 0}\dd(\omega^{\ve_2}f(\ve_2)+f(-\ve_2)) \no\\
+ \frac{1}{\ve_1^2}\le \tilde{\Delta}_2(\ve_1) - \tilde{\Delta}_2(0)\ri.
\end{eqnarray}

By construction, this quantity does not have any singularity in the variable
$\ve_1$.
Constants like $\tilde{\Delta}_2(0)$ and
$\lim_{\ve_2 \rar 0}\dd(\omega^{\ve_2}f(\ve_2)+f(-\ve_2))$ are multiplied by
negative powers of $\ve_1$ and should disappear in Eq.~(\ref{Delta-3-limit-2})
at the end.
We conclude from Eq.~(\ref{Delta-2-limit-2}) that the second term in
Eq.~(\ref{Delta-3-limit-2}) is canceled by the corresponding term in
$\tilde{\Delta}_2(\ve_1)$ and from Eq.~(\ref{Delta-3-limit-2}) that 
\begin{eqnarray} 
\Delta_3(0) = \lim_{\ve_2 \rar 0, \ve_1 \rar 0} J^{-1}\Delta_3(\ve_1,\ve_2,\ve_3)  =  \lim_{\ve_1 \rar 0} \left[\frac{1}{2\ve_1^2} \omega^{-\ve_1} \du^2 (\omega^{\ve_1}f(\ve_1)+f(-\ve_1)) \right.
\nonumber\\ 
\left. + \frac{1}{\ve_1^3} \omega^{-\ve_1} \du \le \omega^{\ve_1}f(\ve_1) + f(-\ve_1) \ri - \frac{1}{\ve_1^3}\lim_{\ve_2 \rar 0}\dd(\omega^{\ve_2}f(\ve_2)+f(-\ve_2)) 
- \frac{1}{\ve_1^2}\tilde{\Delta}_2(0) \right] =
\nonumber\\
\lim_{\ve \rar 0} \frac{1}{4!}\d^2[4\d \omega^{-\ve}\d + 6\omega^{-\ve}\d^2](\omega^{\ve}f(\ve)+f(-\ve)).
\end{eqnarray}
As we can see, the result has a structure similar to
Eq.~(\ref{Delta-2-derivative}), that is a differential operator of a certain
structure acting on $\omega^{\ve}f(\ve)+f(-\ve)$.
We will show that such a structure survives in more complicated cases. 

Repeating for the quantity $\Delta_3(\ve_1,\ve_2,\ve_3)$ the steps that we took
for $\tilde{\Delta}_2(\ve_1,\ve_2,\ve_3)$, we may write 
\begin{eqnarray} \label{lemma3}
\lim_{\ve_2 \rar 0} \dd J^{-1}\Delta_3(\ve_1,\ve_2,\ve_3)  =  \frac{1}{6\ve_1^2}\omega^{-\ve_1} \du^3 (\omega^{\ve_1}f(\ve_1)+f(-\ve_1)) \no\\
+ \frac{1}{\ve_1^5}(\omega^{-\ve_1}f(-\ve_1)+f(\ve_1)-2f(0)) + \frac{1}{\ve_1^4} \lim_{\ve_2 \rar 0}\dd(\omega^{\ve_2}f(\ve_2)+f(-\ve_2)) \no\\
- \frac{1}{2\ve_1^3}\lim_{\ve_2 \rar 0}\dd^2 (\omega^{\ve_2}f(\ve_2)+f(-\ve_2)) \no\\
- \frac{1}{\ve_1^3}\le \tilde{\Delta}_2(\ve_1) - \tilde{\Delta}_2(0)\ri  - \frac{1}{\ve_1^2}\lim_{\ve_2 \rar 0}\dd\tilde{\Delta}_2(\ve_2).
\end{eqnarray}
For ${\Delta}_4$, we thus obtain
\begin{eqnarray}
\Delta_4(\ve_1,\ve_2,\ve_3)  =  \frac{1}{\ve_1\ve_2} \le \Delta_3(\ve_1,\ve_2,\ve_3) - J\Delta_3(\ve_1)\ri 
+ \frac{1}{\ve_1(\ve_1+\ve_2)}\le J\Delta_3(\ve_1) - J\Delta_3(\ve_2)\ri.
\end{eqnarray}
Taking Eq.~(\ref{lemma3}) into account, we may write the analogue of
Eq.~(\ref{Delta-3-limit-2}) for $\Delta_3$ as
\begin{eqnarray} \label{lemma4}
\Delta_4(\ve_1) = \lim_{\ve_2 \rar 0}J^{-1} \Delta_4(\ve_1,\ve_2,\ve_3)  =  \frac{1}{6\ve_1^3}\omega^{-\ve_1} \du^3 (\omega^{\ve_1}f(\ve_1)+f(-\ve_1)) \no\\
+ \frac{1}{\ve_1^6}(\omega^{-\ve_1}f(-\ve_1)+f(\ve_1)-2f(0)) + \frac{1}{\ve_1^5} \lim_{\ve_2 \rar 0}\dd(\omega^{\ve_2}f(\ve_2)+f(-\ve_2)) \no\\
- \frac{1}{2\ve_1^4}\lim_{\ve_2 \rar 0}\dd^2 (\omega^{\ve_2}f(\ve_2)+f(-\ve_2)) \no\\
- \frac{1}{\ve_1^4}\le \tilde{\Delta}_2(\ve_1) - \tilde{\Delta}_2(0)\ri  - \frac{1}{\ve_1^3}\lim_{\ve_2 \rar 0}\dd\tilde{\Delta}_2(\ve_2) 
+ \frac{1}{\ve_1^2}\le \Delta_3(\ve_1) - \Delta_3(0)\ri.
\end{eqnarray}
By construction, this quantity does not have any singularity in the variable
$\ve_1$.
Constants like $\tilde{\Delta}_2(0)$, $\Delta_3(0)$,
$\lim_{\ve_2 \rar 0}\dd(\omega^{\ve_2}f(\ve_2)+f(-\ve_2))$, and
$\lim_{\ve_2 \rar 0}\dd^2(\omega^{\ve_2}f(\ve_2)+f(-\ve_2))$ are multiplied by
negative powers of $\ve_1$ and should finally disappear in Eq.~(\ref{lemma4}).
From Eq.~(\ref{Delta-3-limit-2}), we conclude that the second term on the
r.h.s.\ of Eq.~(\ref{lemma4}) is canceled by the corresponding term in
$\Delta_3(\ve_1)$, while the first term in the last line of Eq.~(\ref{lemma4}) 
is canceled by another term in $\Delta_3(\ve_1)$.
We conclude from Eq.~(\ref{lemma4}) that 
\begin{eqnarray} 
\Delta_4(0) = \lim_{\ve_2 \rar 0, \ve_1 \rar 0} J^{-1}\Delta_4(\ve_1,\ve_2,\ve_3)  =  \lim_{\ve_1 \rar 0} \left[\frac{1}{6\ve_1^3} \omega^{-\ve_1} \du^3 (\omega^{\ve_1}f(\ve_1)+f(-\ve_1)) \right.
\nonumber\\ 
\left. + \frac{1}{2\ve_1^4} \omega^{-\ve_1} \du^2 \le \omega^{\ve_1}f(\ve_1) + f(-\ve_1) \ri - \frac{1}{2\ve_1^4}\lim_{\ve_2 \rar 0}\dd^2 (\omega^{\ve_2}f(\ve_2)+f(-\ve_2))  \right.
\nonumber\\
\left. - \frac{1}{\ve_1^3}\lim_{\ve_2 \rar 0}\dd\tilde{\Delta}_2(\ve_2) - \frac{1}{\ve_1^2}\Delta_3(0) \right] =
\nonumber\\
= \lim_{\ve \rar 0}\frac{1}{6!}\d^3(15\d \omega^{-\ve}\d^2 + 20\omega^{-\ve}\d^3)\M.  
\end{eqnarray}

We may proceed further to higher number $n$ and find the following relations:   
\begin{eqnarray}
\tilde{\Delta}_2(0)   &=&       \lim_{\ve \rar 0} \frac{1}{2!}\d\left[C_2^0 \d \omega^{-\ve}\d^0+C_2^1 \omega^{-\ve}\d\right]\M\no\\
\Delta_3(0)           &=&       \lim_{\ve \rar 0} \frac{1}{4!}\d^2\left[C_4^1 \d \omega^{-\ve}\d + C_4^2 \omega^{-\ve}\d^2\right](\omega^{\ve}f(\ve)+f(-\ve))\no\\
\Delta_4(0)           &=&        \lim_{\ve \rar 0} \frac{1}{6!}\d^3\left[C_6^2 \d \omega^{-\ve}\d^2 +C_6^3\omega^{-\ve}\d^3\right]\M.
\end{eqnarray}
The result for arbitrary number $n>4$ is 
\begin{eqnarray} \label{Delta-n}
\Delta_n(0)           =         \lim_{\ve \rar 0} \frac{1}{(2(n-1))!}\d^{n-1}\left[C_{2(n-1)}^{n-2}\d \omega^{-\ve}\d^{n-2}  + C_{2(n-1)}^{n-1}\omega^{-\ve}\d^{n-1} \right] \M. 
\end{eqnarray}

\section{Reproducing UD functions}

The formula in Eq.~(\ref{Delta-n}) presents the integrands for the two-fold MB
transforms of the UD functions, which we have been searching for.
We recover the dependence on $u$ and $v$ of the function $f(\ve)$ omitted
below Eq.~(\ref{f}) and cast Eq.~(\ref{Delta-n}) in the form
\begin{eqnarray} \label{FR}
\lim_{\ve_2 \rar 0, \ve_1 \rar 0}  \oint_C du~dv~x^u~y^v M_n^{(u,v)}(\ve_1,\ve_2,\ve_3) = 
 \lim_{\ve \rar 0} \frac{1}{(2(n-1))!}\d^{n-1}\Bigl[C_{2(n-1)}^{n-2}\d \omega^{-\ve}\d^{n-2}  \Bigr.\no\\
\Bigl. + C_{2(n-1)}^{n-1}\omega^{-\ve}\d^{n-1} \Bigr] 
\frac{1}{2}\oint_C du~dv~x^u~y^v\left[(x/y)^{\ve} a(\ve)\G(-u-\ve)\G(-v+\ve)  \right. \no\\ 
\left. +  a(-\ve)\G(-u+\ve)\G(-v-\ve) \right]\G(-u)\G(-v) \G^2(1+u+v). 
\end{eqnarray}
In order to demonstrate that the result is correct, we perform the MB
transformations with the integrands for the two-fold MB transforms in
Eq.~(\ref{FR}) and find
\begin{eqnarray}
\frac{1}{2}\oint_C du~dv~x^u~y^v\left[(x/y)^{\ve} a(\ve)\G(-u-\ve)\G(-v+\ve)  \right. \no\\ 
\left. +  a(-\ve)\G(-u+\ve)\G(-v-\ve) \right]\G(-u)\G(-v) \G^2(1+u+v) = 
\nonumber\\
\frac{1}{2} \left[\omega^{\ve} \oint_C du~dv~x^u~y^v D^{(u,v)}[1-\ve,1+\ve,1] + \oint_C du~dv~x^u~y^v D^{(u,v)}[1+\ve,1-\ve,1]\right] = 
\nonumber\\
\frac{1}{2} \left[\omega^{\ve} x^{-\ve}\oint_C du~dv~x^u~y^v D^{(u,v)}[1-\ve] + y^{-\ve}\oint_C du~dv~x^u~y^v D^{(u,v)}[1-\ve]\right] = 
\nonumber\\
y^{-\ve} \oint_C du~dv~x^u~y^v D^{(u,v)}[1-\ve]  =  y^{-\ve} (p^2_3)^{1-\ve}J(1,1,1-\ve).
\end{eqnarray}
where we have adopted the notation of Ref.~\cite{Allendes:2012mr}.
At this point, we use the representation of Ref.~\cite{Usyukina:1992jd} for
$J(1,1,1-\ve)$, which is 
\begin{eqnarray}
J(1,1,1-\ve) = -\frac{1}{(p^2_3)^{1-\ve}}\frac{1}{\ve}\int_0^1d\xi\frac{(y\xi)^{\ve} - (x/\xi)^{\ve}}{y\xi^2 + (1-x-y)\xi + x}.
\end{eqnarray}
Thus, Eq.~(\ref{FR}) may be also written as 
\begin{eqnarray} \label{pro} 
\Phi^{(n)}(x,y) =  \lim_{\ve_2 \rar 0, \ve_1 \rar 0}  \oint_C du~dv~x^u~y^v M_n^{(u,v)}(\ve_1,\ve_2,\ve_3) = \no\\
\lim_{\ve \rar 0} \frac{1}{(2(n-1))!}\d^{n-1}\Bigl[C_{2(n-1)}^{n-2}\d \omega^{-\ve}\d^{n-2}  
+ C_{2(n-1)}^{n-1}\omega^{-\ve}\d^{n-1} \Bigr] \frac{1}{\ve}\int_0^1d\xi\frac{\omega^{\ve}\xi^{-\ve} -\xi^{\ve}}{y\xi^2 + (1-x-y)\xi + x}, 
\end{eqnarray}
where $\Phi^{(n)}(x,y)$ are UD functions of
Refs.~\cite{Usyukina:1992jd,Usyukina:1993ch}.
For $n=2$, we obtain in the integrand 
\begin{eqnarray} 
\lim_{\ve \rar 0} \frac{1}{2}\d\Bigl[\d \omega^{-\ve} + 2\omega^{-\ve}\d \Bigr] \frac{1}{\ve} \Bigl[\omega^{\ve}\xi^{-\ve} -\xi^{\ve} \Bigr] = -\le \ln^3\xi - \frac{3}{2}\ln\omega\ln^2\xi 
+ \frac{1}{2}\ln^2\omega\ln\xi \ri. 
\end{eqnarray}
This coincides with Eq.~(32) of Ref.~\cite{Usyukina:1992jd}.
For arbitrary value of $n$, we may write for the integrand in Eq.~(\ref{pro})
\begin{eqnarray} 
\lim_{\ve \rar 0} \frac{1}{(2(n-1))!}\d^{n-1}\Bigl[C_{2(n-1)}^{n-2}\d \omega^{-\ve}\d^{n-2}  
+ C_{2(n-1)}^{n-1}\omega^{-\ve}\d^{n-1} \Bigr] \frac{1}{\ve} \Bigl[\omega^{\ve}\xi^{-\ve} -\xi^{\ve} \Bigr]  = 
\nonumber\\
- \frac{1}{n!(n-1)!} \ln^{n-1}\xi \le \ln\xi - \ln\omega \ri^{n-1}  \le 2\ln\xi -\ln\omega\ri.
\end{eqnarray}
This result coincides with Eq.~(21) of Ref.~\cite{Usyukina:1993ch}.
The corresponding integrals in Eq.~(\ref{pro}) for arbitrary value of $n$ were
calculated in Ref.~\cite{Usyukina:1993ch}, and the expressions in terms
of simple polylogarithms $\rm{Li}_n$ with composite functions of $x$ and $y$ as
arguments may be found there.
A representation of the integrals in Eq.~(\ref{pro}) that is simpler than the
one of Ref.~\cite{Usyukina:1993ch} was found in Ref.~\cite{Isaev:2003tk},
namely
\begin{eqnarray}
\Phi^{(n)}(x,y) = -\frac{1}{z-\bar{z}}f^{(n)}\le\frac{z}{z-1},\frac{\bar{z}}{\bar{z}-1}\ri,
\end{eqnarray}
where
\begin{eqnarray}
f^{(n)}(z,\bar{z}) = \sum_{r=0}^n\frac{(-1)^r(2n-r)!}{r!n!(n-r)!}\ln^r(z\bar{z})
\le{\rm Li}_{2n-r}(z) - {\rm Li}_{2n-r}(\bar{z})\ri, 
\end{eqnarray}
with the new variables
\begin{eqnarray}
x = z\bar{z},\qquad y = (1-z)(1-\bar{z}).
\end{eqnarray}

This representation may be rewritten in terms of single-valued harmonic
polylogarithms, which were studied in Ref.~\cite{Brown}.
The single-valued harmonic polylogarithms are constructed from harmonic
polylogarithms of Ref.~\cite{Remiddi:1999ew} in such a way that they do not
have cuts in the complex planes of their arguments \cite{Brown}, that is, they
are single-valued in the whole complex plane.
The result is given in Ref.~\cite{Drummond:2013nda} as 
\begin{eqnarray}
f^{(n)}(z,\bar{z}) =  (-1)^{n+1}2L_{\underbrace{0,...,0}_{n-1},\dis{0,1},\underbrace{0,...,0}_{n-1}} (z),
\end{eqnarray}
where the function $L_{0,...,0,0,1,0,...,0} (z)$ is a single-valued harmonic
polylogarithm defined in Ref.~\cite{Brown}.
This result was obtained in Ref.~\cite{Drummond:2012bg} by using symbols,
i.e.\ tensor products of rational functions, which were introduced by
Goncharov in Ref.~\cite{Goncharov:2009} and were applied for the first time to
calculations in quantum field theory in Ref.~\cite{Goncharov:2010jf}.
These symbols correspond to canonical differential $n$-forms, which, upon
integration for admissible pairs of simplices, yield Aomoto $n$-logarithms,
from which Grassmannian polylogarithms may be obtained applying a procedure of
skew-symmetrization \cite{Goncharov:2009,Goncharov:1998kja}.

\section{Conclusion}

The explicit form of the coefficients found in the present paper allows us, on
the one hand, to analytically resum all the ladder diagrams that make up the
solution to the Bethe-Salpeter equation for the case in which the function
$f$ is chosen to be the Euler $\G$ function as in
Ref.~\cite{Broadhurst:2010ds}, and, on the other hand, to write in explicit
form the integration formulae derived in Ref.~\cite{Allendes:2012mr}.
All the values $\Delta_n(0)$ were found in the form of differential
operators acting on the first term of the chain of MB transforms.  
It is plausible that, for other types of functions $f$, different from the
Euler $\G$ function, these recursive relations may be mapped to recursive
relations of other integrable systems in quantum mechanics or condensed-matter
theory.
We conclude that the higher UD functions may be obtained via the application
of a certain differential operator to a simple generalization of the first UD
function, which is the main result of this paper.

\subsection*{Acknowledgments}

The work of B.A.K. was supported in part by the German Science Foundation (DFG)
within the Collaborative Research Center SFB 676 ``Particles, Strings and the
Early Universe'' and by the German Federal Ministry for Education and Research
(BMBF) through Grant No.\ 05H12GUE.
The work of I.K. was supported in part by Fondecyt (Chile) Grants Nos.\
1040368, 1050512, and 1121030,
by DIUBB (Chile) Grant Nos.\ 121909 GI/C-UBB and 125009,
and by Universidad del B\'\i o-B\'\i o and Ministerio de Educacion (Chile)
within Project No.\ MECESUP UBB0704-PD018.
He is grateful to the Physics Faculty of Bielefeld University for accepting
him as a visiting scientist and for the kind hospitality and the excellent
working conditions.
The work of E.A.N.C. was supported in part by Direcci\'on de Investigaci\'on
de la Universidad de La Serena (DIULS) through Grant No.\ PR~12152.
The work of I.P.F. was supported in part by Fondecyt (Chile) Grant No.\
1121030.
The work of M.R.M. was supported in part by Project No.\ MTM2012-32325,
by Ministerio de Ciencia e Innovaci\'on, Espa\~na,
by Fondecyt (Chile) Grant Nos.\ 1080628 and 1120260,
and by DIUBB (Chile) Grant No.\ 121909 GI/C-UBB.

\end{document}